\documentstyle[crna12,12pt]{article}                                            
\begin{document}
\input{psfig}
\begin{flushright}
Tel Aviv U. preprint TAUP-2444-97\\
March 1998                                                               
\end{flushright}
                                                                                
\bigskip                                                                        
                                                                                
\begin{center}{\Large\bf  Kaon Radiative Decay $K^{\pm}  \to \mu \nu \gamma$\\ 
at CKM at the Fermilab Main Injector}                                                   
\end{center}                                                                    

\begin{center}{\large\bf                    
C. Milst\'ene$^a$, P. S. Cooper$^b$, M. A. Moinester$^c$\\    
}\end{center}
\begin{center}\bf{
$^a$ On leave of absence at FNAL from TAU,\\ 
 }\end{center}\begin{center}\bf{caroline@fnal.gov}
\end{center}
\begin{center}\bf{
$^b$ Fermi National Accelerator Laboratory,\\
 }\end{center}\begin{center}\bf{pcooper@fnal.gov }
\end{center}
\begin{center}\bf{
$^c$Tel Aviv University,\\
 }\end{center}\begin{center}\bf{murraym@silly.tau.ac.il}
\end{center}\bigskip\bigskip\bigskip

\begin{center}\bf\it{
Presented by C. Milst\'ene at the Main Injector Workshop,\\Fermilab,
May 1997.\\
}\end{center}\bigskip
\bigskip
\begin{center}{\Large\bf Abstract}\end{center}
\normalsize
High statistics data for the  $K^{\pm}  \to \mu \nu \gamma$ decay can allow
the precision determination of the kaon structure dependent form factors. This
study is possible at the proposed FNAL CKM experiment. CKM (Charged Kaons at
the Main injector) is a decay-in-flight spectrometer with two Ring Imaging
detectors(RICH), and designed to run at the main injector with a high kaon
flux. The radiative decay data might be taken complementary to the primary CKM
effort to study the rare kaon decay ($K \rightarrow \pi \nu \bar{\nu}$) for
which CKM was primarily designed. We summarize here the underlying physics of
$K^{\pm}  \to \mu^{\pm} \nu \gamma$, the experimental status, and how the CKM
spectrometer may be adapted for this study.
\bigskip

\section*{I)~Introduction}

Data on the radiative decay $K^{\pm} \to \mu^{\pm} \nu \gamma$ can give
important information on the properties of the hadronic weak currents of the
K$^{\pm}$ meson. Chiral perturbation theory can be tested in such a process,
since it is characterized by an effective chiral Lagrangian calculable in a
low energy perturbative expansion. The amplitudes to O(p$^4$) order for
radiative semileptonic K decays can be determined from $K^{\pm} \to \mu^{\pm}
\nu \gamma$ decay data.

The principal term of the decay is the Internal-Bremsstrahlung (IB) of charged
particles,  which is completely described by QED. The second term is the
direct emission of the $\gamma$ from the intermediate kaonic states; this term
is structure dependent (SD). The direct emission of the right-handed and
left-handed photon are mediated respectively by vector hadronic currents
dependent on the vector form factor (F$_V$), and axial-vector hadronic
currents dependent on the axial vector form factor (F$_A$). The cross-section
of the structure dependent terms was first derived by Neville \cite{NEVIL} as
a function of these form factors. Since then, many calculations have given
predictions on the values of the form factors. The values obtained  are $\mid
F_V \mid M_K$=0.06 to 0.8 and F$_V$/F$_A$=-0.97 to 0.58
\cite{SARKER,KUMAJ,ROCKMO,SMOES,CARSCHU,MILWAD,DEPOM,PAVSCAD,AVAKY,GASLEU,DON,
AMET,BIEG}.
 
 Chiral Perturbation theory predicts to order O(p$^4$) that the form factor
ratio is given in terms of the combination of chiral coefficients ($L_9^r +
L_{10}^r$), with F$_V$/F$_A$=32$\pi^2 (L_9^r + L_{10}^r)$. Also, the kaon
electric polarizability to order O(p$^4$) is expressed as $\alpha_K = (4
\alpha_f/m_K F_{K}^2)*(L_9^r + L_{10}^r)$, where $\alpha_f$ is the fine
structure constant and $F_{K}$ is the Kaon decay constant. Therefore the 
polarizability is proportional to the ratio F$_V$/F$_A$, which may be measured
in Kaon radiative decay experiments \cite{DONHOL,BABMOIN}.

\begin{figure}[htb]
\centerline{\psfig{figure=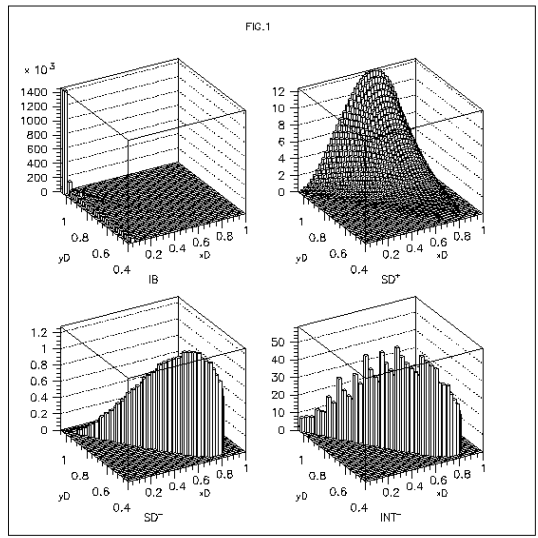,height=10cm}}
\caption { The Dalitz plot for each of the terms contributiong to
the $K^{\pm}  \to \mu \nu \gamma$ decay}
\label{IBSDI}
\end{figure}

  The experimental measurements are generally given in terms of \mbox{$ \mid F_V
-F_A \mid M_K$} for \mbox{SD$^-$} and the interference term (INT$^-$), and 
\mbox{$\mid F_V+F_A\mid M_K$} for \mbox{SD$^+$} and the interference term 
\mbox{INT$^+$}. These terms
populate differently the Dalitz plot representing this three-body decay. The
Dalitz plot population is very sensitive to the phase and size of the
Structure Dependent terms (SD$^{\pm}$), (INT$^{\pm}$), as shown in 
Figure~\ref{IBSDI}  Fig. 1.

It would have been sufficient to measure the interference term in the $K^{\pm}
\to \mu^{\pm} \nu \gamma$ decay, but too few events are available for this
decay. From a measurement of  SD$^{\pm}$ alone, one only gets an ambiguous
evaluation of F$_V$ and F$_A$. In CKM, we plan to measure both the SD$^-$ term
and the INT$^-$ term.

 We will describe the physics motivation in section II, the status of
measurements in section III, FNAL E761 measurement attempts in section IV, the
CKM possibilities in section V, and our conclusions in section VI.

\section*{II)~Physics Motivation}

\begin{figure}[htb]
\centerline{\psfig{figure=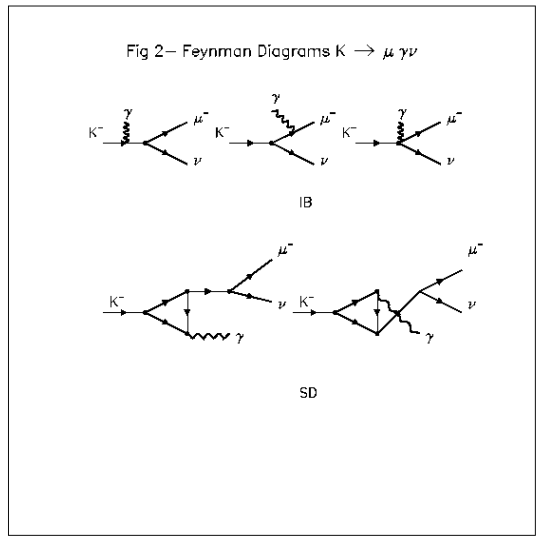,height=10cm}}
\caption{Feynman diagrams of the IB terms and the triangular diagrams of
SD terms }
\label{Feynm}
\end{figure}

  There are three types of contribution to this kaon radiative decay. They are
the Internal-Bremsstrahlung (IB), the structure dependent terms (SD) where the
photon is emitted from intermediate Kaon hadronic states, and the interference
terms between them (INT). The Feynman diagrams of the IB terms and the
triangular diagrams of SD terms are represented in Figure~\ref{Feynm} Fig. 2.

   In the SD terms, the radiative decay $K^{\pm} \to \mu^{\pm} \nu \gamma$
depends on the properties of the hadronic weak currents. There are no hadrons
involved in the final state. Therefore, the kinematics of the three almost
massless point-like particles emitted by the K$^{\pm}$ probes the internal
structure of the kaon without final state interactions.

   The amplitudes SD can be written in terms of F$_V$ and F$_A$. There are two
componants, SD$^+$ proportional to $\mid F_V+F_A{\mid}^2$, and SD$^-$
proportional to $\mid F_V-F_A{\mid}^2$. The SD$^+$ and SD$^-$ terms correspond
to the emission of photons with positive and negative helicities, respectively.
The interference term is composed of two terms, INT$^+$ and INT$^-$ which
correspond to the interference of IB with SD$^+$ and SD$^-$ respectively
\cite{AKIBA,NIK}. The interference terms allow a determination of the sign of
the form factors, and thereby the sign of the amplitudes. The Dalitz plot
expresses decay probability as a function of x (related to the $\gamma$
energy) and y (related to the muon energy). Figure~\ref{IBSDI} Fig. 1 shows that the different
terms populate different regions of the Dalitz plot.

The decay rate is then given as:\\
   $$d^2\Gamma/dxdy = \Gamma(IB) + \Gamma (SD^+) + \Gamma (SD^-)+
      \Gamma (INT^+)+\Gamma (INT^-). $$ 

We have:\\
   $$\Gamma(IB) =A_{IB} \bullet f_{IB}(x,y)$$
   $$\Gamma(SD^+) = A_{SD}\bullet f_{SD}^+ (x,y)
    \bullet ((F_V+F_A)\bullet M)^2 $$ 
$$\Gamma(SD^-) = A_{SD}\bullet f_{SD}^- (x,y)\bullet((F_V-F_A)*M)^2 $$  
$$\Gamma(INT^+)= A_{INT}\bullet f_{INT}^+ (x,y)\bullet ((F_V+F_A) \bullet M)$$
$$\Gamma (INT^-)= A_{INT}\bullet f_{INT}^- 
(x,y)\bullet ((F_V-F_A)\bullet M).$$

Here, 
A$_{IB}$ = W$_{K\mu}^2 \bullet \frac{\alpha_f}{2\pi}\bullet\frac{1}{(1-r)^2}$ 
                                                                
       A$_{SD}$ = A$_{IB}\bullet \frac{1}{4r}\bullet \frac{M^{4}_{K}}{F_K^2}$
   
       A$_{INT}$ = A$_{IB}\bullet \frac{M_K^2}{F_K}$

W$_{K\mu}^2$ =0.6351 is the probabitlity for the decay 
$K^- \to \mu^- \nu$,\\ 

$x=\frac{2E^{CM}_{\gamma}}{M_K}$, $y =\frac{2E^{CM}_{\mu}}{M_K }$,
$r=\frac{M_{\mu}^2}{M_K^2}$\\

    $ f_{IB}(x,y)=\left(\frac{1-y +r} {x^2 \bullet 
(x+y-1-r)}\right)\bullet \left(x^2 +2\bullet(1-x)\bullet(1-r)-
    \frac{2\bullet x\bullet r\bullet(1-r)}{x+y-1-r}\right)$\\

    $ f_{SD}^+ (x,y)= [x+y-1-r]\bullet 
       [(x+y-1)\bullet (1-x)-r] $\\

   $ f_{SD}^- (x,y)= [1-y +r]\bullet [(1-x)(1-y)+r] $\\

   $ f_{INT}^+ (x,y)=\left(\frac{1-y+r} {x\bullet (x+y-1-r)}
\right)\bullet[(1-x)\bullet (1-x-y)+r]$\\

  $ f_{INT}^- (x,y)=\left(\frac{1-y+r} {x\bullet (x+y-1-r)}
\right)\bullet [x^2 -(1-x)\bullet (1-x -y)-r]$.\\ 

\newpage
\section*{III)~Experimental Status}

 Previous experiments with measurements of F$_A$ and F$_V$ are reported in
Table~\ref{tabpi01} together with the predicitons from Chiral 
Perturbation theory
\cite{DON,BIEG}. The theory value quoted for\mbox{ $\mid F_V+F_A{\mid}M_K$} is
obtained using the $K^-  \to  e \nu \gamma$ decay data.

\begin{table}[h]
\begin{tabular}{|c|c|c|}            \hline
                &         &       \\
Predictions from&$\mid F_V+F_A{\mid} M_K$=0.134&$\mid F_V-F_A{\mid}M_K$=0.049\\  
Chiral Theory   &          &      \\  \hline\hline
                &          &       \\
CERN PS         &$\mid F_V +F_A{\mid}_{e}M_K$=&-2.5$<\mid F_V -F_A{\mid}_{e}M_K<$
0.5 \\
Experiment      &0.153 $\pm$0.01&     \\  \hline
                  &         &       \\
KEK             &$\mid F_V +F_A{\mid}_{\mu}M_K<$0.23 & -2.5$<\mid F_V -
F_A{\mid}_{\mu}M_K<$0.3 \\
Experiment      &          &     \\  \hline
                  &         &       \\
ITEP PS& -1.2$<\mid F_V +F_A{\mid}_{\mu}M_K<$1.1&
-2.2$<\mid F_V -F_A{\mid}_{\mu}M_K<$0.6 \\
Experiment        &          &     \\  \hline
&         &       \\
BNL 787        &$\mid F_V +F_A{\mid}_{\mu}M_K$& Not measured \\
Experiment    &= 0.165$\pm$0.007$\pm$0.011&     \\  \hline
\end{tabular}
\caption{Theoretical Predictions and Experimental status}
\label{tabpi01}
\end{table}

 The Cern PS experiment $K^-  \to e \nu \gamma$ was performed utilizing a
stopped K- separated beam in a counter experiment, with a lead-glass detector
for the $\gamma$  \cite{CERN}. They had 51 $\pm$ 3 events to determine $SD^+$
and 9 events to determine limits on $SD^-$. The KEK experiment $K^-  \to \mu
\nu \gamma$ used a High Resolution Spectrometer with a NaI(T1) $\gamma$
detector and a stopped K- separated beam \cite{AKIBA}. In the \mbox{SD$^+$} 
region 3.44 events were obtained after fit, while  in the \mbox{SD$^-$+INT$^-$}
region, 142 events were measured. The ITEP-PS experiment 
$K^- \to \mu^- \nu \gamma$, used a
700-liter Xenon Bubble Chamber and stopped Kaon from a separated beam, with
442 $K^-  \to \nu \mu \gamma$ candidates \cite{DEMI}. E761 attempted to
measure \mbox{$\mid F_V -F_A{\mid}M_K$}, as described in E761 reports
\cite{NIK,FNAL2}. The most recent value of \mbox{$\mid F_V+F_A{\mid}M_K$} was
measured with good precision using 2693 $K^-  \to \mu^- \nu \gamma$ obtained at
BNL with stopped kaons. Their result confirms the $K^-  \to e^-\nu \gamma$
experimental value, (and universality), and they both confirm the predictions
from Chiral theory \cite{DON,AMET,BIEG}. They were also able to measure for
the first time the Branching ratio BR(SD$^+$) of the order of 10$^{-5}$ with a
precision of the order of 20$\%$ \cite{CONVERY,CONVER2}. CKM will be sensitive
to the \mbox{SD$^-$} term, and should be able to get a precise measurement, including
the sign of \mbox{$(F_V-F_A)~M_K$}.

To summarize the present situation, a value was obtained from previous CERN
and BNL data for $\mid F_V+F_A{\mid}M_K \approx 0.153 - 0.165$. For 
\mbox{$\mid F_V-F_A{\mid}M_K$}, the best limits from the different 
experiments are in the range [-2.2,+0.3] \cite{AKIBA,DEMI}.

\section*{IV)~ E761 Studies}
 
The 375 GeV/c Hyperon unseparated beam with 5\% K$^-$ and the E761 High
Resolution Spectrometer in the cascade setup were used. The background to the
$K^-  \to \nu \mu \gamma$ channel comes from K decays with $\pi^0$. These
decays have larger branching ratio than the $\gamma$ channel, with $\pi^0 \to
\gamma \gamma$ faking the $\nu \gamma$, e.g.
$$~K^- \to  \pi^- \pi^0  \ \         ;\ \      BR = (21.17 \pm 0.05)\%$$
$$~K^- \to  \pi^- \pi^0 \pi^0 \ \ ; \ \    BR = (1.73 \pm 0.04)\% $$
$$~K^- \to \mu \nu \pi^0 \ \       ; \ \     BR = (3.18 \pm 0.08)\%$$
$$~K^- \to e \nu \pi^0  \ \        ; \ \    BR = (0.55 \pm 0.028)\%.$$

Backgrounds which come from the other Hyperons in the beam; (e.g., the decays
$\bar{\Sigma}^- \to \bar{p} \pi^0$, $\Xi^- \to \Lambda \pi^-$) were removed by
the analysis cuts. These cuts also removed part of the background coming from
competing K$^-$ decays, e.g. $K^{-}\to e^{-} \nu \pi^{0}$ decay \cite{FNAL2}.
Other sources of background were the interactions of the beam in detector
materials.

 An E761 analysis \cite {FNAL2} considered a partial sample of 2838 events.
Taking into account the energy measured in the PbG detector and the position
of the highest energy cluster, the components of the photon momentum were
computed. The momentum resolution of the lead glass detector was ($\sigma
P_\gamma$/$P_\gamma$=.033+.3276/$\sqrt E_\gamma$) and the angle precision was
$\sigma \theta_\gamma$=0.66 mrad. The missing mass square to the neutrino was
then computed, using the measured momentum of the incoming and charged
outgoing particles, and also the momentum of the $\gamma$ computed from the
information in the PbG detector. The momentum resolutions of the incoming and
the charged outgoing particle are better than for the $\gamma$ and are
reported in the table in the next paragraph. The particles are not identified.
 
\begin{figure}[htb]
\centerline{\psfig{figure=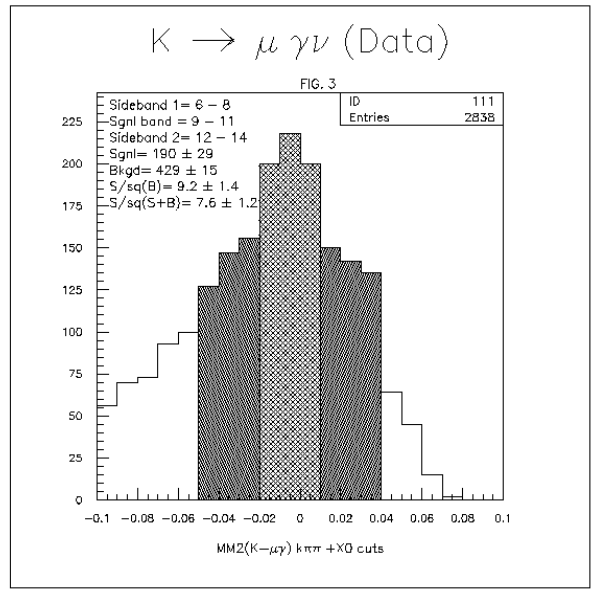,height=11.cm}}
\caption {Side-bands background substraction in the square of the missin+
mass MM2(K-$\mu \gamma$) }
\label{MM2NU}
\end{figure}

 Figure~\ref{MM2NU} Fig. 3 shows the square of the missing-mass, where the peak is rather wide
due to the poor resolution in mass which comes from the poor resolution in the
momenta used to calculate it. 

In Figure~\ref{MM2NU} Fig. 3, one can see 2838 events. Considering
a side-band background substraction, this can be separated into $\simeq
190\pm$ 29 events located in a peak at 0 GeV, above a background of 429 $\pm$
15 events.

The above summary shows that the signal to background ratio in E761 was not
good enough to get quality data on the Kaon radiative decay. This is so
because of the absence of $\mu$ identification, because of the limited $\gamma$
detector resolution, and because the background from decays and interactions
were very high compared to the signal.

\section*{V)~ Kaon radiative decay with CKM}

\indent
1) We expect a 30 MHZ Kaon separated beam at the main injector. 
This will lead to:
 $ K^+ \  Decay \ \ \ \ \ \               6MHZ $ 
                
 $ BR(K \mu \nu\gamma)\ \ \ \ \  5. \bullet 10^-3$
               
 $ Acceptance\ \ \                    1.\bullet10^-2$

 $ \ \ \ \ \ \ \ \ \ \ \ \ \ \ \ \ \ \     \longrightarrow
\longrightarrow\longrightarrow$

 $ \ \ \ \ \ \ \ \ \ \ \ \ \ \ \ \ \ \ \        300\  HZ  $
      
 $ \ \ \ \ \ \ \ \ \ \ \ \ \ \ \ \ \ \ \         10^5$  sec/week

 $ \ \ \ \ \ \ \ \ \ \ \ \ \ \ \ \ \ \ \      \longrightarrow\longrightarrow
\longrightarrow\longrightarrow\longrightarrow\longrightarrow\longrightarrow$

 $ \ \ \ \ \ \ \ \ \ \ \ \ \ \ \ \ \ \ \       30$ Million  events/week  

 Here we define 1 week = 100 beam-hours * 1000 sec/hour. Therefore, in one
week, we expect the number of $ K^- \to \nu \mu \gamma$ decays 
to be higher than all previous experiments together.

\begin{table}[ht]
\begin{tabular}{|c|c|c|}            \hline
                  &         &       \\
                  & E761    & CKM   \\  
                  & DATA    & MC  \\  \hline
$\sigma P_K$/$P_K$&0.77$\%$&0.24$\%$ \\
                  &          &     \\  \hline
$\sigma P_\mu$/$P_\mu$&1.70$\%$&0.78$\%$ \\
                      &          &     \\  \hline
$\sigma P_\gamma$/$P_\gamma$&.033+.3276/$\sqrt E_\gamma$&
0.01/$\sqrt E_\gamma$ \\
                          &          &     \\  \hline
$\sigma \theta_\mu$       &28 $\mu$rad&124 $\mu$rad \\
                          &          &     \\  \hline
$\sigma \theta_\gamma$    &1.52 mrad-3.34*$P_\gamma/P_K$& 
0.66 mrad\\
                          &          &     \\  \hline                                
\end{tabular}
\caption{Resolution Comparison}
\label{tabpi02}
\end{table}

\newpage

\begin{figure}[htb]
\centerline{\psfig{figure=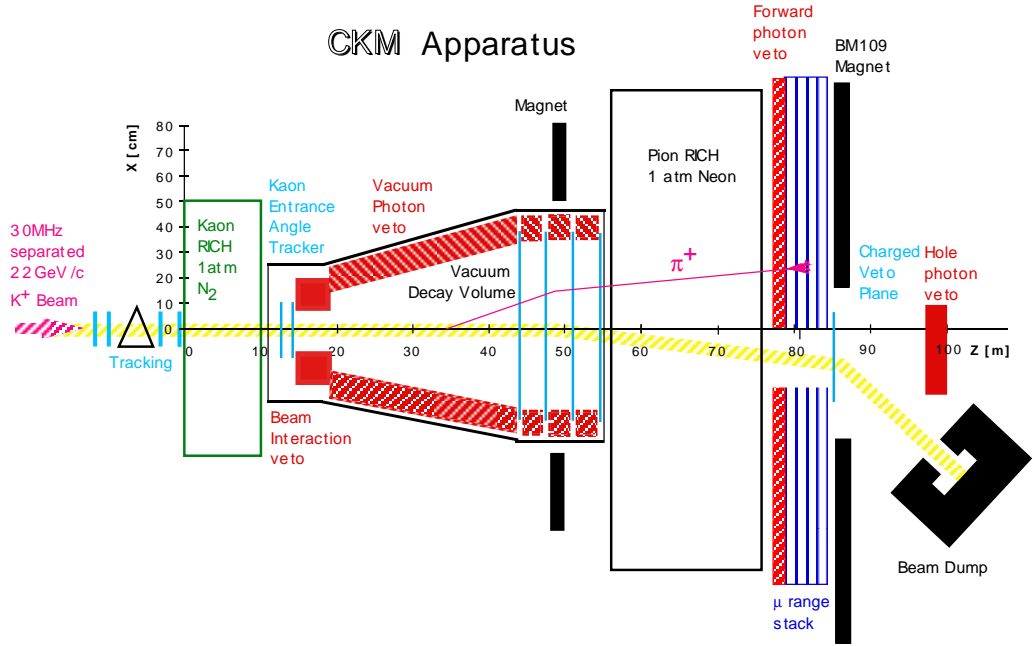,height=10.cm}}
\caption {The CKM spectrometer }
\label {APPAR}
\end{figure}

 2) The CKM apparatus is shown in Figure~\ref{APPAR} Fig. 4. CKM \cite{PETE} is built with two
RICH detectors. 
 
One is a RICH detector ten meters long at the threshold of the
22 GeV kaon beam; containing CF4 at 0.7 atm pressure. The second RICH
is twenty meters long at the pion threshold; containing Neon at 1 atmosphere.
Both RICH detectors are separated by a vacuum decay volume 30 meters long and
covered by a  $\gamma$ veto layer. A PbG stack at the rear of the second RICH
is an additional veto to clean the background. The PbG stack is followed by a
$\mu$ range stack and a bending magnet to tilt the beam to the beam dump. An
additional PbG stack located eighty meters from the entrance of the first RICH
covers the central hole.

  We have simulated the CKM system with a CsI $\gamma$ detector of the KTEV 
variety added at the rear of the pion RICH and before the PbG veto stack, at about sixty meters
from the entrance of the Kaon RICH. This $\gamma$ detector will be 15 cm wide
in the y direction and 70 cm wide in the x direction, and shifted by 10 cm
from the center of the beam. Each of the 75 blocks of CsI will be 4x4 cm$^2$,
as may be seen in Figure~\ref{APPAR}Fig. 4. In the CsI detector, a nitrogen laser similar to
that used by KTEV is needed to calibrate the digital readout system
\cite{KTEV}. This detector system also allows kaon identification in the Kaon
RICH, as well as muon identification using the pion RICH and the $\mu$ range
stack. Such information were not avalaible in E761. Together with the photon
veto, the PbG stack, and the better resolution photon detector, CKM will
achieve both a better identification of the events of interest and a better
background rejection.

 We compare the expected photon resolution of the CsI detector to the photon
resolution achieved in E761. Those resolutions are obtained using the EXP
Monte Carlo. The comparison is made with E761 data. Most of the comparative
resolutions are reported below in Table~\ref{tabpi02}.

  3) We achieved with the E761 PbG detectors the following resolutions across
the Dalitz plot: $$\sigma x \simeq 0.3\ E761\ \ ;\   \sigma x \simeq 0.015\ \
CKM$$
$$\sigma y \simeq 0.05\ E761\ \ ;\   \sigma y \simeq 0.003\ \ CKM$$  

\begin{figure}[htb]
\centerline{\psfig{figure=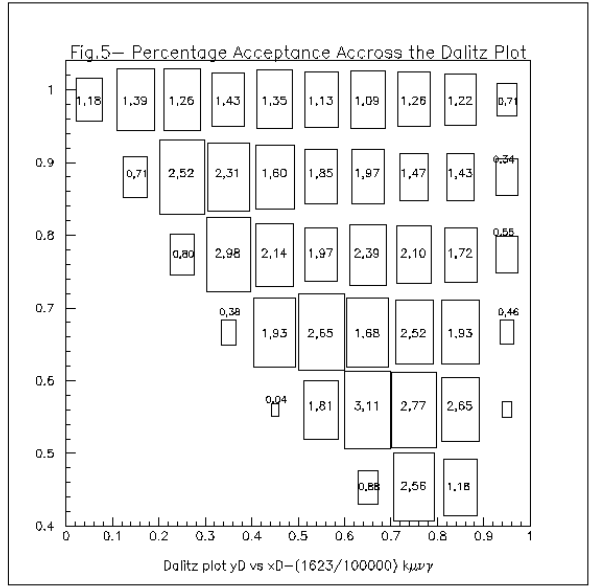,height=11.cm}}
\caption {Percentage Acceptance accross the Dalitz plot,a total of 1623 +
 out of 100000 events were seen}
\label {DACCEPT}
\end{figure}

The acceptance reached using the EXP \footnote{EXP is a fast Monte-Carlo written
by P.S.Cooper which had had many lifes starting with a BNL/FNAL experiment and
through e761 and e781 went over a new transformation for CKM}   
is of the order of 1\% as used in the evaluation of the number of events one 
obtains in a week of CKM run. A plot of the Acceptance per bin across the 
Dalitz plot is represented in Figure~\ref{DACCEPT} Fig. 5.
 
In that figure, the bin size is twice 
the resolution in x and y respectively (a 2$\sigma$ margin was allowed). We generated 100,000 $K^-  \to \nu \mu \gamma$
decays with EXP-MC. In Fig. 5, we show the percentage Acceptance across the
Dalitz plot.

\section*{VI)~Conclusion} 

  In the Main injector, a week of CKM running will provide more $K^-  \to \nu
\mu \gamma$ than all previous experiments together. The background will be
minimized, considering the vacuum in the decay volume, the photon veto, and
the PbG stacks. The events will be better identified, having both kaon and muon
identification. The signal will be better measured for all the particles,
especially the photons.

\section*{Acknowledgements} Special thanks are due to G. Burdmann, C. Quigg, 
for interesting discussions. Thanks are due to S. Kananov and
A. Ocherashvili for help with the GEANT simulations of the E761 studies. This
work was supported in part by the U.S.-Israel Binational Science Foundation,
(BSF),  Jerusalem, Israel; and in part by A. Zaks to whom one of us is 
particularly thankful.   
\listoftables
\listoffigures
                                                                      

\begin{thebibliography}{99}                                          
\bibitem{NEVIL} D. E. Neville, Phys. Rev. 124, 2037, (1961).
\bibitem{SARKER} A. Q. Sarker, Phys. Rev. 173, 1749 (1968). 
\bibitem{KUMAJ} W. K. Kummer, W. Majerotto, Nuovo Cimento 55, 558 (1968).
\bibitem{ROCKMO} R. Rockmore, Phys. Rev. 177, 2573 (1969).
\bibitem{SMOES}  M. G. Smoes, Nucl. Phys. B20, 237 (1970).
\bibitem{CARSCHU}N. J. Carron, R. L. Schult, Phys. Rev. D1, 3171 (1970).   
\bibitem{MILWAD} K. A. Milton, W. W. Wada, Phys. Lett. 98B, 367 (1980).
\bibitem{DEPOM} D. A. Bryman, P. Depommier, C. Leroy, Phys. Reports
88 (1982) 151.
\bibitem{PAVSCAD}M. Pavers and M. D. Scadron, Nuovo Cimento 78A, 159 (1983).
\bibitem{AVAKY}  E. Z. Avakyan{\it et al.}, Dubna Preprint R2-86-441, (1986).
\bibitem{GASLEU} J. Gasser and H. Leutwyler, Nucl. Phys. B250, 517 (1985). 
\bibitem{DON}  J.F. Donoghue, Preprint UMHEP-329-1990
\bibitem{AMET} Ametller,{\it et al.}, Phys. Lett. B303, 140 (1993).
\bibitem{BIEG} J. Bijnens, G. Ecker, J. Gasser, Nucl. Phys. B396, 87 (1993).
\bibitem{DONHOL} J. F. Donoghue, B. R. Holstein, Phys. Rev. D40, 2378 (1989)
\bibitem{BABMOIN} D. Babusci, S. Belluci, G. Giordano, G. Matone, A. M. 
Sandorfi, M. A. Moinester, Phys. Lett. B277, 158 (1992), and references therein.
\bibitem{AKIBA} Y. Akiba {\it et al.}, Phys. Rev. D32, 2911 (1985).  
\bibitem{NIK} V. S. Demidov, M. A. Kubantsev, A. N. Nikitenko, 
ITEP Preprint ITEP-30-1993, E761 FNAL Report HNOTE 524.
\bibitem{MOINEST} M. A. Moinester, Proceeding of the Workshop on Virtual Compton
Scattering,Clermont-Ferrand,France,Edit.V.Breton,June1996,TelAviv Univ.Preprint,
TAUP-2354-96, and references therein.  
\bibitem{CERN} J. Heintze {\it et al.}, Nucl. Phys. B149, 365  (1979) 
\bibitem{DEMI} V. S. Demidov {\it et al.}, Sov. J. Nucl. Phys. 52, 1006 (1990) 
\bibitem{CONVERY} M. R. Convery, Princeton University, UMI-97-07305,
Ph.D. Thesis (1996).
\bibitem{CONVER2} M. R. Convery, Proceeding of the 1996 Minneapolis
DPF Conference.
\bibitem{FNAL2} P. S. Cooper and C. Milst\'ene, E761 FNAL Report HNOTE 685,
V.S. Demidov, M.A.Kubantsev and A.N. Nikitenko,Preprint ITEP-30-1993,
HNOTE 524. 
\bibitem{PETE} P. S. Cooper, Contribution to the Main Injector Workshop,
Fermilab, May 1997.
\bibitem{KTEV} R. S. Kessler {\it et al.}, Nucl. Inst. Meth. A368, 653 (1996) 
\end{thebibliography}
\end{document}